\documentstyle[12pt,aps]{revtex}
\begin{document}
\def\ApJ{{\sl Astrophys. J.}}
\def\la{\mathrel{\mathpalette\fun <}}
\def\ga{\mathrel{\mathpalette\fun >}}
\def\fun#1#2{\lower3.6pt\vbox{\baselineskip0pt\lineskip.9pt
  \ialign{$\mathsurround=0pt#1\hfil##\hfil$\crcr#2\crcr\sim\crcr}}}
\title{\ \\\ \\\ \\\ \\\ \\
Possible Systematic Decreases in the Age of Globular Clusters}
\author{X. Shi$^a$, D. N. Schramm$^{a,b}$, D. S. P. Dearborn$^c$
and J. W. Truran$^a$\\
\vskip 0.3cm
$^a$The University of Chicago, Chicago, IL 60637-1433\\
$^b$NASA/Fermilab Astrophysics Center, Fermi National Accelerator Laboratory,
Box 500, Batavia, IL 60510-500\\
$^c$Lawrence Livermore National Laboratory, Livermore, CA 94550}
\vskip 2.8cm
\maketitle

\centerline{\bf Abstract}
The ages of globular clusters inferred from observations
depends sensitively on assumptions like the initial helium abundance and the
mass loss rate. A high helium abundance (e.g., $Y\approx$0.28) or a
mass loss rate of $\sim$10$^{-11}M_\odot$ yr$^{-1}$
near the main sequence turn-off region
lowers the current age estimate from 14 Gyr to about 10--12 Gyr,
significantly relaxing the constraints on the
Hubble constant, allowing values as high as 60km/sec/Mpc for
a universe with the critical density and 90km/sec/Mpc for a baryon-only
universe. Possible mechanisms for the helium enhancement in globular
clusters are discussed, as are arguments for
an instability strip induced mass loss near the turn-off.
Ages lower than 10 Gyr are not possible even with the operation
of both of these mechanisms unless the initial helium abundance in globular
clusters is $>0.30$, which would conflict with indirect measurements of helium
abundances in globular clusters.
\vskip 0.45in
\noindent ------------------------------------------------

\noindent Submitted to {\sl Science}
\vfill\eject
\section{Introduction}
The latest estimate of the age of globular clusters (GCs)
is about 14 Gyr with an spread of different clusters
of about $\pm$2 Gyr \cite{Sandage93a}.
This severely constrains $H_0$, the Hubble constant, for a $\Lambda$
(the vacuum energy density)
=0 Friedman-Robertson-Walker cosmology.
For $\Omega$ (the ratio of the density of the universe to the
critical density)$=1$, the Hubble constant $H_0$
must be less than 45km/sec/Mpc \cite{Sandage93a}; while
for a baryonic universe where $0.01/h^2<\Omega<0.015/h^2$
(with $h=H_0/100$km/sec/Mpc) as constrained
by Big Bang Nucleosynthesis \cite{Walker1991},
$H_0\la 70$km/sec/Mpc. Currently
dynamic measurements of the density of the universe favor
$\Omega>0.3$ \cite{NusserDekel1993}, which yields $H_0\la 58$ km/sec/Mpc.
Direct measurements of the Hubble constant, however, yield values that
range widely from 40km/sec/Mpc to 100 km/sec/Mpc \cite{Huchra1992}.

A reconcilation of the direct
measurement of the Hubble constant and that inferred
from the age of the globular clusters may well require not only improvements
of the distance scale systematics but also
continued investigation of the assumptions implicit in the
inferred age of globular clusters.

It is noted that the estimate of the age depends sensitively on many
uncertainties of globular clusters,
such as the actual turn-off luminosity (distance) determination,
the metallicity, the oxygen to iron ratio, the helium diffusion in stars,
and the initial helium abundance. To this we add the uncertainty due to
an unobservable low mass loss rate inferred from lithium observations of stars.
The uncertainties of the turn-off luminosity, the metallicity and
the oxygen to iron ratio have been widely discussed in the
literature \cite{Wheeler89,V92,Bergbusch1992},
and has been included in the current age estimate of $14\pm 2$
Gyr \cite{Sandage93a}. The helium diffusion in stars is also estimated to
lower the age of GCs by $5\%$--15$\%$
\cite{ProffittMichaud1991,ProffittVandenberg1991,Chaboyer92}.
Uncertainties in the initial helium abundance and the mass loss rate of
stars near the turn-off point, however, have not been thoroughly discussed,
and a full consideration of them may significantly influence the age estimate.
\section{Age dating of globular clusters}
Two methods are commonly used to estimate the age of a globular cluster
(see review of ref. \cite{Demarque91}).
One is to fit theoretical isochrones (temperature-luminosity
curves for stars at certain ages)
to the shape of a globular cluster on the observed color-magnitude diagram,
using a distance modulus determined from fitting either a zero age main
sequence (ZAMS) or a zero age horizontal branch (ZAHB).
It is also necessary to transform
a luminosity-temperature diagram of a theoretical calculation
to a color-magnitude diagram with an interstellar reddening
correction \cite{Bergbusch1992}.
Besides of the uncertainties mentioned in the introduction,
this method is very sensitive to the transformation between the
luminosity-temperature diagram and the color-magnitude diagram,
the fudicial sequence of globular clusters extracted from their scattered
color-magnitude diagram, the subjectivity in fitting isochrones to the
fiducial sequence, and the effectiveness
of a constant mixing length in describing the convection.
It also depends sensitively on the adopted
interstellar reddening if using ZAMS fitting to obtain distances.

The difficulty in improving the fitting method lies mainly in the fact that
the color of both observed GCs and theoretical isochrones on the
color-magnitude diagram is hard to determine precisely.
The color of observed GCs depends on the
interstellar reddening which can be very uncertain;
while the color of isochrones is sensitive to the treatment of convection
in stellar models. Figure 1 shows isochrones with similar parameters but
different mixing lengths (parametrized by $\alpha$, the ratio of
the mixing length to the scale height of pressure) in the convection zone.
The color shift due to a different choice of
$\alpha$ is especially prominent on the upper subgiant and the red giant branch
of isochrones. Although the uncertainty in choosing different
mixing lengths may be partially resolved by calibrations of ZAMS
stars \cite{Vandenberg1988},
it still persists since mixing lengths may change in different
evolution phases, such as the subgiant branch and the red giant branch.

A more reliable method is to relate the
age with the turn-off luminosity of a GC, regardless of its
detailed morphology. The turn-off luminosity is obtained
from the calibrated visual magnitude of RR Lyrae stars on the
horizontal branch (HB) and the observed visual magnitude differences
between RR Lyrae stars and main-sequence turn-off points, with bolometric
corrections \cite{Sandage93a,IbenRenzini1984}
(i.e., using a distance from ZAHB fitting).
This method is independent of the interstellar reddening,
the subjectivity of fitting isochrones to the observed
GC sequence and the mixing length adopted in stellar modeling.
It also suffers from the uncertainties mentioned in the introduction.

\section{The Helium Uncertainty}
The age turn-off luminosity relation of GCs has been
shown by Iben and Renzini \cite{IbenRenzini1984} to be
\begin{equation}
\log \Bigl({{\rm Age}\over {\rm 1 yr}}\Bigr)
=8.497-1.88(Y-0.24)-1.44(Y_{\rm HB}-Y)-0.088\log Z
+0.41\Delta M^{\rm RR}_{\rm TO}{\rm (bol)},
\end{equation}
where $Y$ is the helium abundance of GCs, $Z$ is their metallicity,
$Y_{\rm HB}$ is the helium abundance at the envelope
of the HB stars, and $\Delta M^{\rm RR}_{\rm TO}$(bol) is the
bolometric magnitude difference between the
RR Lyrae stars and the turn-off point.
The coefficients in the equation have been refined over the years, but no
significant changes have been made.

$Y_{\rm HB}-Y$ is estimated to be 0.01--0.02 \cite{IbenRenzini1984}.
Therefore a strong dependence of the age on the helium abundance
$Y$ is manifest in eq. (1).
The $Y$ in GCs cannot be directly measured because the helium
line strength is weak at the temperature involved.
Instead $Y$ is indirectly inferred from comparisons between
stellar models and observations. One way is to compare the observed
color and period of stars at the blue edge of the instability strip
with the helium dependent temperature and pulsation period relation
found in stellar models. This method
yields 0.20$<Y<0.30$ \cite{IbenRenzini1984,Bingham1984}.
However, as mentioned above, the temperature
of stars cannot be confidantly inferred from their observed colors due to
the uncertainty in the interstellar reddening and stellar modelling.

The latest efforts compare the measured ratio of the number of stars
on the horizontal branch to the number of stars
with higher luminosities on the red giant branch
with that predicted by stellar
models \cite{Buzzoni1985,CMP1987,Bencivenni1989}.
These efforts yield $Y$ values for individual clusters
that scatter between 0.20 and 0.30, with an average of 0.24 and
a spread of $\pm 0.02$.
However, this method may suffer from uncertainties in treating
the semi-convection and mass loss, both of which affect
the life-time of the HB phase \cite{Castellani1971}.
A recent comparison between the morphology of the horizontal branch in 47 Tuc
and stellar models also suggested
0.2$<Y<0.3$ \cite{DormanVandenbergLaskarides1989}.
So far, all calculations constrain the helium abundances in
GCs to be less than 0.30.

A maximal helium abundance of 0.30 in GCs seems also to be consistent with that
inferred from luminosities of the HB stars.
It is well known that the bolometric magnitude of RR Lyrae stars,
$M_{\rm RR}{\rm (bol)}$, is
sensitive to $Y_{\rm HB}$ at their envelopes \cite{IbenRenzini1984}:
\begin{equation}
M_{\rm RR}{\rm (bol)}=0.943-3.5(Y_{\rm HB}-0.30)+0.183\log Z.
\end{equation}
Current calibrations of
the visual magnitude of RR Lyrae stars yield a 0.2--0.3 magnitude spread
for different methods, and an uncertainty of
0.2 magnitude within each method \cite{SandageCacciari1990,Sandage1993b}.
Therefore, an uncertainty of
0.06 in $Y_{\rm HB}$ (or equivalently $Y$ since $Y_{\rm HB}-Y\approx
0.01$) is marginally tolerable by determinations of
luminosities of the horizontal branch. In other words, $Y$ can be as high
as 0.30 for individual clusters, given a primordial value of 0.24.
It should be noted that the observation of RR Lyrae stars
in the disk of the galaxy may not serve a valid constraint on
$Y_{\rm HB}$ in globular clusters.

It is conventionally assumed that $Y=0.24$
(as in ref. \cite{Sandage93a}) or 0.235
(as in ref. \cite{Bergbusch1992}) in GCs, in agreement with the observed
primordial helium abundance $Y=0.235\pm 0.01$
\cite{Skillman1993}
and the prediction of Big Bang Nucleosynthesis
\cite{Walker1991}.
However, as discussed above, a higher $Y$ is allowed, that
can significantly reduce the age estimate of GCs.
For example, a $Y$ of 0.3 will reduce the age by as much as 30$\%$
with respect to a $Y$ of 0.24, according to eq. (1).

Determining the age of GCs with the isochrone fitting method has a similar
$Y$ sensitivity as in eq. (1) if a distance modulus from ZAHB fitting is
used.  If a distance from ZAMS fitting is used in the isochrone fitting,
the effect of a higher Y in GCs on their age is not easy to quantify.
Figure 2 shows 10 Gyr and 11 Gyr isochrones with $Y=0.30$ and $Z=0.0002$,
compared with isochrones of 12 Gyr, 14 Gyr and 16 Gyr with
$Y=0.24$ and the same $Z$.
The figure also shows the 11 Gyr $Y=0.30$ isochrone
shifted by $\Delta (B-V)$=0.05 and $\Delta M{\rm (bol)}$
=0.18 magnitude. Therefore, if we use the
magnitude of ZAMS as a distance indicator, a 11 Gyr $Y=0.30$
isochrone may resemble a 14 Gyr $Y=0.24$ isochrone at the main sequence
and the subgiant branch, if a reddening of 0.05 is added.
The mismatch at the red giant branch may be easily remedied by a modest
adjustment of the mixing length as seen from Fig. 1.

A helium abundance higher than the primordial value
in GCs is possible to achieve under certain scenarios.
The simplest scenario is the enrichment by the first generation
supernovae (SNe) of GCs \cite{BrownBurkertTruran1991}.
Measurements of low metallicity HII regions in
irregular galaxies and extrogalactic HII regions showed that
their helium abundance $Y$ correlates
with their oxygen abundance $O$ by $Y=0.24+130O$
(i.e., $dY/dZ=5$) \cite{Skillman1993,Pagel1992}.
For GCs with [Fe/H]=$-1$, their oxygen abundances are about 1/3 of the
solar value \cite{Wheeler89}, i.e., $O\approx 2\times 10^{-4}$. If GCs follow
the same helium-oxygen correlation as those low metallicity HII regions,
then $Y\approx 0.27$, which is significantly above the primordial value.
Of course, it is debatable whether GCs follow the same $Y$-$O$
relation as low metallicity HII regions,
although both of them have low metallicities compared with the solar value.
The sun, on the other hand, has $Y=0.28$ and
$O=6\times 10^{-4}$, which would require a different relation
$Y=0.24+70O$ (i.e., $dY/dZ=3$).
Therefore, solar-type disk stars must have had quite a different chemical
history from low metallicity gases, like those of GCs and HII regions.
It is worth noting that VandenBerg, Bolte and
Stetson \cite{VandenbergBolteStetson1990} found
a 2--3 Gyr age spread among GCs with [Fe/H]$>-1.6$ and no significant age
spread among GCs with [Fe/H]$<-1.6$. If an enhancement in the helium abundance
correlates with an increase of metallicity in GCs, as suggested above,
the age spread among metal-rich GCs may be easily
at least in part explained by different enhancements
of the helium abundance in different clusters.

Exotic models, like Pop III stars
\cite{CarrBondArnett1984,BondArnettCarr1984},
may enrich helium but not heavier elements. It is calculated that
if the oxygen cores of these Pop III stars are larger than $\sim 100M_\odot$,
the cores will collapse completely into blackholes
without ejecting heavy elements.
However, processed helium in Pop III stars can be deposited into the
interstellar medium through mass loss and mixing of Pop III stars.
The maximal amount of helium enrichment is $\Delta Y=0.17$ for an initial
helium abundance of 0.24, if all gases are processed once in Pop III stars.
Therefore, in order to enhance the helium abundance in a $10^6M_\odot$
cloud from 0.24 to about 0.30, at least $3\times
10^5M_\odot$ gas has to be turned into Pop III stars.
This requires quite a different mass function from the
$M^\alpha$ power law with $\alpha=-2.35$
observed today \cite{Salpeter1955}. For the metallicity of the
cloud is low, the required mass function must have $\alpha>-0.5$, so that
there are no more than $\sim 100$ SNe in a $10^6M_\odot$ cloud.

An enhancement of the helium abundance may also arise during the
formation of globular clusters. For example,
the difference in the first ionization energy between the
hydrogen and the helium (13.6eV vs. 24.6eV) enables the
helium to recombine earlier than the hydrogen,
and hence prone to earlier collapses.

One class of model for forming globular clusters, based on Peebles \&
Dicke \cite{PeeblesDicke1968},
argues that GCs form from primordial baryon density
fluctuations after recombination, which have a Jeans mass similar to
the mass of typical GCs, 10$^6 M_\odot$.
According to the standard recombination picture of the universe, hydrogen
recombined at $T\sim 3000$K (redshift $z\sim 1000$) \cite{Peebles1993}
while helium recombined at $T\sim 10^4$K
or $z\sim 3000$ (inferred from Fig. 5 of ref. \cite{GouldThakur1970}).
In the period after helium recombination and before hydrogen recombination,
the helium falls quickly into the potential well of dark matter,
following the density fluctuation of dark matter \cite{KolbTurner1990}.
As hydrogen recombines,
the density fluctuations in the hydrogen and the helium begin to grow
jointly. The overdensed region, however, already has an enhanced helium
abundance from the in-fall of helium early on. Assuming the density fluctuation
in dark matter at $z\sim 1000$ to be $\delta\rho/\rho$ and the primordial
helium abundance $Y_{\rm p}=0.24$, the enhanced helium abundance
in the overdense region at $z\la 1000$ is
\begin{equation}
Y={Y_{\rm p}(1+\delta\rho/\rho)\over Y_{\rm p}(1+\delta\rho/\rho)
+1-Y_{\rm p}}\approx 0.24(1+0.76\delta\rho/\rho).
\end{equation}
To enhance the helium abundance to $Y=0.30$ on the 10$^6M_\odot$ scale
requires a $\delta\rho/\rho\sim 1/3$ on this scale,
which means $\delta\rho/\rho$ on the 10$^6M_\odot$ scale
will become nonlinear at a $z$ of several hundred. This cannot be realized
in the standard cold dark matter (CDM) model, where structures
at this scale become non-linear
at $z\la {\cal O} (10)$ \cite{KolbTurner1990}, but may be possible
in models where primordial seeds such as topological defects
provide nonlinear fluctuations on small scales
\cite{Vilenkin1985,TurokSpergel1991,LuoSchramm1993}.
A $\delta\rho/\rho$ of 1/3 at $z\sim 1000$ with $\sim$10$^6M_\odot$ scales
will certainly distort the cosmic microwave background radiation (CMBR)
and produce non-gaussian fluctuations at the arcsecond angular scale in CMBR.
Potentially conflicting CMBR anisotropies reported on the 0.5 to 3$^\circ$
angular scales are interpreted by some as possibly being indicative of
non-gaussian fluctuations \cite{Turok1993,Schramm1993,Luo1994}.

Another class of models, based on that of Fall and Rees \cite{Fall1985},
argues that GCs form out of cold clouds with $\sim 10^6M_\odot$ in
proto-galaxies \cite{MathewsSchramm1993}.
A higher helium content may then arise if Pop II stars form in the shock
waves from the first generation supernovae in the cloud
\cite{BrownBurkertTruran1993}. The temperature
of these clouds is typically 10$^4$K due to sharp drop in the cooling
function of the gas cloud at 10$^4$K \cite{GouldThakur1970}.
At this temperature hydrogen is partially ionized and interacts with
the radiation fields in the cloud, but helium remains neutral.
The sound speed in helium, $c_{\rm He}$, is then $\sim 10$ km/sec,
while the sound speed in hydrogen, $c_{\rm H}$, is much larger.
(The reason for different
sound speeds is that perturbations in hydrogen can propagate much faster
by interactions between hydrogen and photons, while perturbations in helium
have to propagate at least initially through interactions between helium atoms
and ambient atoms or charged particles.) When the velocity of a
shock wave drops below $c_{\rm H}$, the shock no longer compresses
hydrogen in the cloud medium into the shock front but continues to do so with
helium. Therefore stars formed from the compressed gas at the shock front will
have a higher helium abundance than the primordial value.

To have all the gas in the 10$^6M_\odot$ cloud
swept by a $\sim$10km/sec shock, the energy output of SNe has to be
$0.5(10^6M_\odot){c_{\rm He}}^2\approx 10^{51}$ ergs without considering
energy losses of the shock, which is the kinetic energy output of one
SN. If we consider the radiation and ionization losses of the shock,
the energy output of SNe has to be at least 10 times larger, or
$\ga 10^{52}$ergs \cite{BrownBurkertTruran1991}.
By requiring the shock to sweep through the cloud
within $10^7$yrs, the formation timescale of $\sim 1M_\odot$ stars,
we need a SN rate of $\ga 10^{-6}$ yr$^{-1}$ per 10$^6M_\odot$, which
is roughly in line with the SN rate required to generate a metallicity
of $10^{-1}$--$10^{-2}Z_\odot$ in GCs \cite{BrownBurkertTruran1991},
and is also consistent with the order of magnitude estimate of SN rates
($\sim 10^{-2}$ per year per galaxy) inferred from the quasar absorption
line systems \cite{LauroeschTruranWeltyYork1993}.

The propagation of the shock front is illustrated in Figure 3. After the
speed of the shock drops below $c_{\rm H}$, the helium abundance in the
compressed shell of the shock front is
\begin{equation}
Y\approx{Y_{\rm p} M_{\rm sh}\over (1-Y_{\rm p}) M_{\rm tot}\delta/l+Y_{\rm p}
M_{\rm sh}}
\end{equation}
where $M_{\rm sh}$ is the mass swept by the shock, $M_{\rm tot}$ is the total
mass of the cloud, and $Y_{\rm p}\ (=0.24)$ is the helium
abundance before the shock.
$\delta$ is the thickness of the shock front and $l$ is
the thickness of cloud after being caved by the shock wave.
If the SNe rate $\ga 10^{-6}$ yr$^{-1}$ per cloud, the shock wave can
propagate through the entire cloud at a speed larger than $c_{\rm He}$.
Then $M_{\rm sh}\sim M_{\rm tot}$. If $l\gg \delta$, i.e., the density
inside the shell is much larger than the density outside,
$Y$ can be significantly larger than $Y_{\rm p}$.

We have briefly discussed several scenarios to enhance the helium abundance
in GCs. There may be other possible scenarios, such as a magnetic field, that
may also segregate charged gas from neutral gas.
If those scenarios operate only in GCs, they will not have a direct
effect on HII regions where the primordial helium abundance is measured.
The small amount of helium-enriched gas from disrupted GCs can only
contaminate interstellar medium modestly. Helium-enriched stars from
disrupted GCs may also constitute only a small population of halo stars.
Once again, a spread in $Y$ after these helium enhancement processes may
be a candidate to account at least in part for the spread in age
estimates for different GCs which assume a universal $Y$.

\section{The Uncertainty in Mass Loss}
Mass loss has been proposed to explain the lithium depletion
in F type Pop I stars with surface temperatures of 6600$\pm 200$K
and were motivated by the coincidence of their locations
with an extrapolation of the instability strip down to the main
sequence \cite{SchrammSteigmanDearborn1990}.
The same mechanism may also operate in some Pop II stars which are
observed to have lithium depletion. The position of the instability
strip on Pop II main sequence is quite uncertain.
By a simple extrapolation from Pop I stars, the temperature range
of the instability strip is in the vicinity of
the turn-off temperature of GCs \cite{DearbornSchramm1993}.
The narrowness of the observed lithium dip constrains any proposed
mass loss to a restricted temperature range between 6500K and 6700K.

The temperatures of GC stars increase until they eventually turn off and evolve
to the giant branch. Without mass loss, stars spend about 90\%
of their lifetime getting to the turn-off, and about 10\% following the
turn-off. When mass loss occurs at the temperature of the turn-off, the
small reduction in mass causes the model to redden earlier. The
turn-off is then seen to be at a lower luminosity, and stars appear to spend
an increased fraction of their life beyond the turn-off. This
results in bump on the luminosity function. This bump, in fact,
further constrains the magnitude of mass loss that is possible.
The GCs will then appear older due to their lower turn-off luminosities
according to eq. (1)
\cite{DearbornSchramm1993,WilsonBowenStruck-Marcell1987}.
A narrow instability strip does not affect the
lower main sequence and so avoids the arguments against large-scale
mass loss through out the entire lower main
sequence \cite{SchrammSteigmanDearborn1990}.

The mass loss rate needed to explain the lithium depletion in Pop II stars
is about 10$^{-11}M_\odot$ yr$^{-1}$, which is far too small to be
detected directly. It may, however, have detectable impacts on the luminorsity
function of GCs \cite{SwensonFaulkner1992}
due to above mentioned bump in the luminosity function.
Therefore, globular cluster
observations utilizing high-sensitivity CCD photometry may provide a definitive
test of the mass loss hypothesis. Observations of low metallicity blue
stragglers, on the other hand, seems to indicate the instability stripe is
bluer than the turn-off point of GCs with similar metallicities
\cite{Carney1994}. However, blue stragglers may be too pathological to
draw conclusions on the main sequence of GCs.

Dearborn and Schramm \cite{DearbornSchramm1993}
have calculated the effect of a mass loss rate
$\sim 10^{-11}M_\odot$ yr$^{-1}$ in the temperature range of $6500\pm 200$K
and $6600\pm 200$ K on the evolution of GCs. Such assumptions result in GCs of
11 Gyr to 13 Gyr old looking 2$\sim 3$ Gyr older.

To combine the mass loss with a higher helium abundance, we construct
isochrones based on Dearborn's stellar code with a gaussian mass loss rate
\begin{equation}
{\rm Mass\ Loss\ Rate}=10^{-11}M_\odot {\rm yr}^{-1}\exp\Bigl[-({T-6500{\rm K}
\over 240{\rm K}})^2\Bigr],
\end{equation}
and $Y=0.28$. $Z$ is taken to be 0.0002 ([Fe/H]=$-2.0$) for the isochrones.
Figure 4 shows such isochrones at
10 and 11 Gyr old as well as standard isochrones without mass loss and with
$Y=0.24$ at 14 Gyr, 16 Gyr and 18 Gyr old. Again, the 11 Gyr isochrone
with mass loss and a higher helium shifted by $\Delta (B-V)$=0.04
and $\Delta M{\rm (bol)}$=0.14 may resemble standard isochrones at $\sim 16$
Gyr. Quantitatively, the age-$\Delta M_{\rm TO}^{\rm RR}({\rm bol})$
relation for these non-standard isochrones is well approximated by
\begin{equation}
\log\Big({{\rm Age}\over {\rm 1 yr}}\Bigr)=15.41
-3.45\Delta M_{\rm TO}^{\rm RR}({\rm bol})
+0.546\big[\Delta M_{\rm TO}^{\rm RR}({\rm bol})\bigr]^2.
\end{equation}
For [Fe/H]$\approx -2$ clusters such as M30, M68, M92 and
NGC6397, age estimates from eq. (6) will be 2--3 Gyr lower than
the estimates of Sandage using standard assumptions\cite{Sandage93a},
as shown in Table 1.
\section{Summary}
As seen above, either a higher helium abundance of $Y\approx 0.28$ and/or a
mass loss rate of $\sim 10^{-11}$M$_\odot$ yr$^{-1}$ at temperatures around
$6500$--$6600$ K will significantly lower the age of GCs by 2--3 Gyr.
Therefore the currently quoted age estimate of 14$\pm 2$ Gyr is potentially
subject to systematic shifts and its uncertainties are model dependent and
not well represented by the quoted error.
A GC age of 10 Gyr is not out of reach considering the aforementioned
uncertainties, and is entirely consistent with the age of the beginning of
nucleosynthesis of heavy elements (i.e., SNe events) measured by radioactive
dating that gives a lower limit of 10
Gyr \cite{MeyerSchramm1986,CowanThielemannTruran1991}.
It is also consistent with the age of the disk, 10$\pm 2$ Gyr,
measured from the cooling of white dwarfs \cite{Winget1987},
provided that the disk collapsed from the halo within 1 or 2 Gyr.

A GC age lower than 10 Gyr, however, is still untenable even with a
combination of a reasonably high $Y$ and a reasonable
mass loss rate, unless $Y>0.30$ is allowed, which would conflict with
indirect measurements of $Y$ in GCs.

Better determinations of the helium abundance in GCs and the mass loss
hypothesis need improved observations of GCs and understanding of stellar
physics. In particular, the determination of the luminosity functions of
GCs through observation and the improved modelling of
the horizontal branch and red giant branch stars are essential.

If GCs are indeed 10 Gyrs old,
the current constraints on cosmological parameters will be significantly
relaxed. For a baryonic universe where $0\le\Omega\ll 1$,
the Hubble constant can be as high
as 90 km/sec/Mpc; for $\Omega> 0.3$, $H\la 72$km/sec/Mpc;
if $\Omega=1$, $H<60$km/sec/Mpc. Obviously, if the measurements of the Hubble
constant eventually converge, it will put constraints on $\Omega$ in
a $\Lambda=0$ Friedman-Robertson-Walker universe and improve
our understanding of the age of globular clusters.
\section{Acknowledgement}
We thank Bruce Carney, Sangjin Lee, Grant Mathews, Michael Turner and
Don York for helpful discussions. This work
is supported by the DoE (nuclear) and NASA at the University of Chicago, by
the DoE at Livermore, and by the DoE and NASA through grant 2381 at Fermilab.
\vfill\eject

\vfill\eject
\begin{table}
\caption{The age estimates from eq. (6) with a higher helium abundance
and mass loss vs. the age estimates of Sandage [1]
for four globular clusters with [Fe/H]$\approx -2$.
\label{table1}}
\begin{tabular}{cccc}
Globular Clusters& [Fe/H]& Ages from Eq. (6)
& Ages from Sandage [1]\\
M30&$-2.13$& 11.5 Gyr  & 13.7 Gyr\\
M68&$-2.09$& 10 Gyr    & 12.1 Gyr\\
M92&$-2.24$& 13 Gyr    & 15.2 Gyr\\
NGC 6397&$-1.91$& 13 Gyr& 15.4 Gyr\\
\end{tabular}
\end{table}
\vfill\eject
\centerline{\bf Figure Captions}
\noindent Figure 1. Two isochrones with similar parameters but
different mixing lengths. The dashed line: $\alpha=1.50$; the solid line:
$\alpha=1.69$.
\medskip

\noindent Figure 2. Isochrones with and without an enhancement of the
helium abundance in GCs. The solid lines: 12 Gyr, 14 Gyr and 16 Gyr
isochrones with $Y=0.24$ and $Z=0.0002$; the long-dashed lines:
10 Gyr and 11 Gyr isochrones with $Y=0.30$ and the same $Z$;
the short-dashed line: the 11 Gyr $Y=0.30$ isochrone
shifted by $\Delta (B-V)$=0.05 and $\Delta M({\rm bol})$
=0.18 magnitude.
\medskip

\noindent Figure 3. The propagation of a SN shock front inside a 10$^6M_\odot$
cloud.
\medskip

\noindent Figure 4. Isochrones with and without a higher helium abundance
and mass loss. The solid lines: 14 Gyr, 16 Gyr and 18 Gyr isochrones with
$Y=0.24$ and no mass loss; the long-dashed lines:
10 and 11 Gyr isochrones with $Y=0.28$ and a mass loss
rate that satisfies eq. (5); the short-dashed line:
the 11 Gyr isochrone with  $Y=0.28$ and mass loss
shifted by $\Delta (B-V)$=0.04 and $\Delta M({\rm bol})$=0.14.

\end{document}